\documentclass[conference]{IEEEtran}
\IEEEoverridecommandlockouts    

\usepackage{multirow}
\usepackage{lipsum}
\usepackage{amsmath}
\usepackage{amssymb}
\usepackage{graphicx}
\graphicspath{{./Figures/}}

\usepackage{algorithm}
\usepackage{algorithmic}
\usepackage{booktabs}
\usepackage{makecell}
\usepackage{array}
\usepackage{pifont}
\usepackage{xspace}
\usepackage[table]{xcolor}
\newcommand{\method}{ARISE\xspace}
\newcommand{\cmark}{\ding{51}}  
\newcommand{\xmark}{\ding{55}}  
\newcommand{\collisionrate}{collision rate\xspace}
\newcommand{\runredlight}{frequency of running red lights\xspace}
\newcommand{\runstopsign}{frequency of running stop signs\xspace}
\newcommand{\outofroad}{average distance driven out of road\xspace}
\newcommand{\followroute}{route following stability\xspace}
\newcommand{\routecompletion}{average percentage of route completion\xspace}
\newcommand{\timespent}{average time spent to complete the route\xspace}
\newcommand{\acceleration}{average acceleration\xspace}
\newcommand{\yawvelocity}{average yaw velocity\xspace}
\newcommand{\laneinvasion}{frequency of lane invasion\xspace}
\newcommand{\overallscore}{overall score\xspace}

\newcommand{\collisionrateabbr}{CR\xspace}
\newcommand{\runredlightabbr}{RR\xspace}
\newcommand{\runstopsignabbr}{SS\xspace}
\newcommand{\outofroadabbr}{OR\xspace}
\newcommand{\followrouteabbr}{RF\xspace}
\newcommand{\routecompletionabbr}{Comp\xspace}
\newcommand{\timespentabbr}{TS\xspace}
\newcommand{\accelerationabbr}{ACC\xspace}
\newcommand{\yawvelocityabbr}{YV\xspace}
\newcommand{\laneinvasionabbr}{LI\xspace}
\newcommand{\overallscoreabbr}{OS\xspace}

\usepackage{newfloat}
\usepackage{listings}
\usepackage{url}
\usepackage{makecell} 
\usepackage{comment}
\floatstyle{ruled}
\newfloat{listing}{tb}{lst}{}
\floatname{listing}{Listing}

\title{\LARGE \bf ARISE -- \underline{A}daptive \underline{R}efinement and \underline{I}terative \underline{S}cenario \underline{E}ngineering}

\author{Konstantin Poddubnyy$^{1}$, Igor Vozniak$^{1}$, Ivan Burmistrov$^{1}$, Nils Lipp$^{1}$, Davit Hovhannisyan$^{1}$, \\ Christian Müller$^{1}$, Philipp Slusallek$^{1}$
\thanks{$^{1}$German Research Center for Artificial Intelligence (DFKI) GmbH, Campus D3.2, 66123 Saarbruecken, Germany}%
\thanks{$^{2}$\protect\url{https://github.com/kopodfki/ARISE}}%
\thanks{\textbf{Accepted for publication at the IEEE Intelligent Vehicles Symposium (IV), 2026.}}%
}

\begin{document}
	
	\maketitle
	\thispagestyle{empty}
	\pagestyle{empty}
	
	\begin{abstract}
		The effectiveness of collision-free trajectory planners depends on the quality and diversity of training data, especially for rare scenarios. A widely used approach to improve dataset diversity involves generating realistic synthetic traffic scenarios. However, producing such scenarios remains difficult due to the precision required when scripting them manually or generating them in a single pass. Natural language offers a flexible way to describe scenarios, but existing text-to-simulation pipelines often rely on static snippet retrieval, limited grammar, single-pass decoding, or lack robust executability checks. Moreover, they depend heavily on constrained LLM prompting with minimal post-processing.
        To address these limitations, we introduce \method — \textbf{A}daptive \textbf{R}efinement and \textbf{I}terative \textbf{S}cenario \textbf{E}ngineering, a multi-stage tool that converts natural language prompts into executable Scenic scripts through iterative LLM-guided refinement. After each generation, \method tests script executability in simulation software, feeding structured diagnostics back to the LLM until both syntactic and functional requirements are met. This process significantly reduces the need for manual intervention. Through extensive evaluation, \method outperforms the baseline in generating semantically accurate and executable traffic scenarios with greater reliability and robustness.
	\end{abstract}

	\section{Introduction}
    
	\label{sec:introduction}
	Autonomous driving (AD) systems require thorough testing across a wide spectrum of traffic scenarios, including out-of-distribution scenarios, which are either rarely seen in the real-world, or impractical to capture, due to the underlying costs, and danger to be staged in physical experiments. Thus, simulation-based traffic scenario generation has emerged as a practical and cost-efficient alternative to real-world testing. The comprehensive and cost-efficient safety-critical scenarios can potentially be created on demand using the domain-specific tools like Scenic~\cite{fremont2023scenic} and OpenSCENARIO~\cite{openscenario2020}, and thereby expose autonomous vehicle (AV) models to unusual and hazardous situations and conditions.

    Large Language Models (LLMs) have recently gained researchers’ attention for transforming natural language descriptions into executable simulated traffic scenarios. Using the rich knowledge encapsulated in them, LLMs allow for the direct translation of the plain text describing a scenario into the simulation specific code. ChatScene~\cite{zhang2024chatscene} introduced an LLM-based tool that generates driving scenarios in Scenic format, executable within the CARLA simulator (\url{https://carla.org})~\cite{dosovitskiy2017carlaopenurbandriving}, by integrating a structured knowledge base with LLM-generated responses to construct runnable scripts. Numerous tools (described in more detail in the Related Work section) leverage LLMs to transform free-form text descriptions (e.g. traffic accident reports, short prompts) into the Scenic scripts. These works demonstrate the advantages of using LLMs to automate traffic scenario generation.

    \begin{figure}[t]
      \centering
       \includegraphics[width=0.78\linewidth]{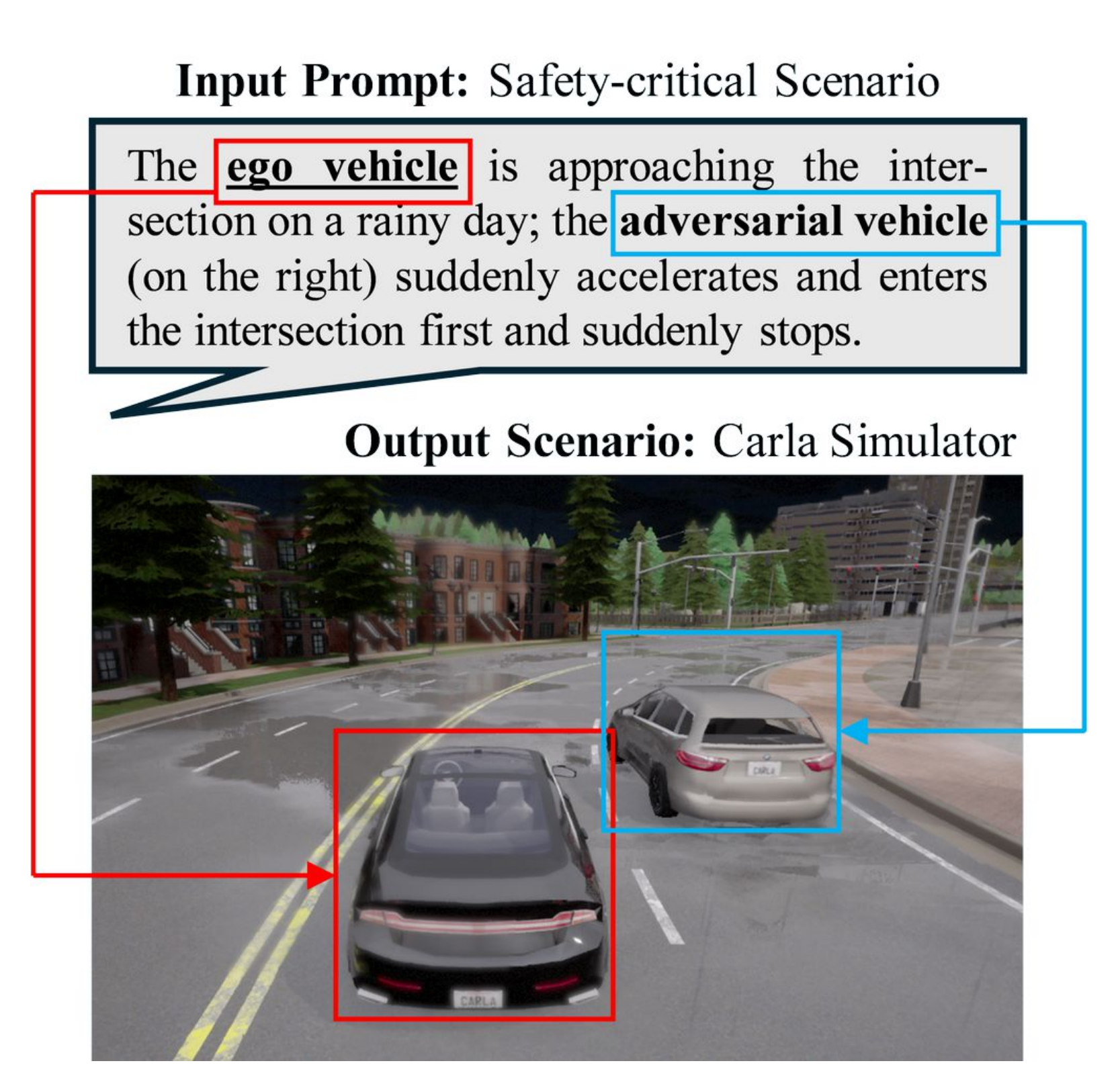}
       \caption{Prompt-based traffic scenario generation (Carla rendered).}
       \label{fig:teaser_short}
    \end{figure}

    Although there is a clear trend toward future scenario generation pipelines that rely on human-language-based precision, the challenge of reliably translating natural language into executable simulation scripts remains unresolved. Compilation, simulation and runtime errors might render a resulting script unusable. Existing works tend not to test the executability of resulting scripts, requiring manual corrections in case of errors. This undermines the intended automation of such pipelines and highlights the need for a more robust and adaptive solution capable of producing reliable and refined results.

    To tackle these challenges, we present \textbf{\method}$^{2}$: \textbf{A}daptive \textbf{R}efinement and \textbf{I}terative \textbf{S}cenario \textbf{E}ngineering, an advanced tool for simulated traffic scenario generation from natural language descriptions (cf. Figure~\ref{fig:teaser_short}). \method directly extends the open-source ChatScene codebase by integrating an important Test-and-Repair Loop (TRL) into the generation pipeline. Similar to the ChatScene, \method initially combines the knowledge base,  prompting and LLM to draft a candidate Scenic script, which is then automatically compiled and tested in the CARLA simulator. When errors are detected, \method automatically initiates an iterative repair process, significantly minimizing the need for user intervention. By feeding the diagnostic feedback (e.g. error stack traces, debug output) from the previous attempts back to the LLM with a structured prompt, the TRL seeks to produce revised and valid output. Thus, \method transforms the single-pass generation into a dynamic dialogue between the generator and the LLM. Leveraging the continuous advancement mechanism, \method offers improved success rates and practical robustness when compared to earlier works. In our experiments, \method achieves higher \textbf{Execution Success Rate (ESR)} \eqref{eq:esr} - the ratio of the resulting traffic scenario scripts that execute without error and without user intervention, to the total number of generated scripts. Additionally, we use the \textbf{Repair Convergence Rate (RCR)} \eqref{eq:rcr}, which indicates how quickly the repair loop converges to a valid scenario, and the \textbf{Semantical Conformity Score (SCS)} \eqref{eq:scs}, which provides insight into the semantical correspondence between the initial scenario descriptions and resulting Scenic scripts. The experiments demonstrate very good results, proving the efficiency of the proposed advancements.

    In summary, our contributions include:
    \begin{itemize}
    	\item \method, an LLM-driven closed-loop traffic scenario generator that introduces a novel iterative feedback and improvement loop, enabling automatic debugging and correction of the Scenic-based simulation scripts.
    	\item Improved performance, where \method outperforms the prior approach in terms of execution success and repair convergence rates (ESR and RCR respectively).
    	\item Robustness and contextual consistency, where we demonstrate that \method delivers results that are not only syntactically correct and executable, but are also semantically conform to the initial descriptions.
    \end{itemize}
	
	\section{Related Work}
	\label{sec:relatedwork}

    \subsection{Synthetic Traffic Scenario Generation}
    
    A number of surveys \cite{f100, ding2023surveysafetycriticaldrivingscenario, schütt20231001waysscenariogeneration} provide comprehensive taxonomies of early approaches to synthetic traffic scenario generation that relied on programmatic or knowledge-based methods rather than LLMs. Non-LLM scenario generation techniques are generally categorized into knowledge-driven, data-driven, and generative approaches.
    Knowledge-driven methods use expert-defined rules and structured scenario representations. Industry efforts have led to formal scenario description languages and ontologies~\cite{winner2019pegasus}, while academic work has explored encoding traffic knowledge through enumeration of scenario components~\cite{klueck:18} and identification of edge cases~\cite{Bogdoll_2023}.
    With the growth of real-world driving datasets, many methods turned toward data-driven techniques. Datasets from projects such as Waymo~\cite{sun2020scalabilityperceptionautonomousdriving} have enabled pipelines like AC3R~\cite{Huynh:19}, which reconstructs scenes from accident reports. Other techniques mine these datasets to uncover and recreate rare or critical driving events \cite{so:19}.
    
    Generative methods synthesize new safety-critical scenarios via probabilistic or reinforcement learning-based techniques. For instance, \cite{Zhao_2017} employ importance sampling to generate rare driving maneuvers, while other works extend this by introducing additional vehicles or perturbing trajectories to explore the boundaries of safe driving behavior~\cite{liu2024safetycriticalscenariogenerationreinforcement}. Recent work also applies real accident data to model causal factors in collisions~\cite{wei:24}. In addition, deep generative models have been introduced to model scenario distributions directly. SceneGen \cite{tan2021scenegenlearninggeneraterealistic} learns to sample realistic but abstract traffic scenes from real data, and TrafficGen~\cite{feng2023trafficgenlearninggeneratediverse} applies a diffusion model to generate diverse traffic configurations conditioned on map context. Complementary methods~\cite{Feng_2021_1, Feng_2021_2, 9204818} produce catalogs of scenarios that systematically test safety constraints.
    These non-LLM methods laid important groundwork for automated scenario generation, introducing structured representations, simulation-ready data pipelines, and optimization-driven scenario synthesis.
    
    \subsection{LLM-empowered Frameworks}
    
    LLMs have recently shown strong potential in traffic scenario generation, with several systems exploring different strategies to combine language understanding with scene synthesis. Some approaches guide generative models to create detailed traffic scenes or visual assets. For instance, CTG++~\cite{zhong2023languageguidedtrafficsimulationscenelevel} introduces a scene-level LLM-guided conditional diffusion model, blending the strengths of diffusion processes with textual conditioning to generate diverse traffic scenes. Similarly, ChatSim~\cite{wei2024editablescenesimulationautonomous} employs LLMs in combination with image and video generators to create photo-realistic assets and scene videos, aiming at high-fidelity visual outputs.
    
    Another category of methods leverages LLMs for structured scene construction and traffic dynamics generation. LCTGen~\cite{tan2023languageconditionedtrafficgeneration} uses a transformer-based decoder in conjunction with an LLM to select maps and initialize traffic distributions and vehicle dynamics. TTSG~\cite{ruan2025trafficscenegenerationnatural} converts natural language inputs into structured JSON scenarios for use in the CARLA driving simulator, facilitating straightforward integration with simulation platforms.
    
    To enhance controllability and scene realism, retrieval-augmented and iterative refinement approaches have also emerged. OmniTester~\cite{l:24} combines an LLM with the SUMO road network generator and Retrieval-Augmented Generation (RAG)~\cite{chen2025omniragcomprehensiveretrievalaugmentedgeneration}, enabling generation of both dynamic elements and static infrastructure from user prompts. Additionally, an LLM-based evaluator is used to iteratively refine outputs, focusing on background vehicles’ behavior.
    
    Other systems focus on improving interpretability and edge-case diversity. LLMScenario~\cite{llmscenario:24} uses Chain-of-Thought (CoT) prompting and evaluation-based feedback to produce plausible yet rare scenarios. Lastly, LeGEND~\cite{tang2024legendtopdownapproachscenario} takes a hierarchical approach: it first translates abstract textual descriptions into logical structures via one LLM, then instantiates specific scenario parameters using a second LLM and search-based algorithms.
    
    Despite their innovations, existing approaches share several key limitations. Many lack robust functional and semantic validation, struggle with ambiguous or novel language inputs, and often produce outputs that are hard to execute directly in simulation environments. Several methods do not include feedback or repair loops, leading to brittle generation pipelines. Others rely heavily on pretrained generators or static world models, which limits domain adaptability, temporal reasoning, and structural consistency in the resulting scenarios.
    
    \subsection{LLM-based Scenario Generation Using Scenic}
    
    Recent works have focused on leveraging LLMs for automatic generation of synthetic traffic scenarios from textual descriptions. ChatScene~\cite{zhang2024chatscene} leverages a combination of an LLM with a knowledge base containing domain examples to produce scripts encoding traffic scenarios in single pass. Given a user prompt input, ChatScene decomposes the scenario into a set of key components: relative spawn positions, adversarial behavior and road geometry. It then retrieves the matching Scenic code snippets from the database, and assembles them into a full script that can be potentially executed in the CARLA simulation software.
    
    In contrast, ScenicNL~\cite{elmaaroufi2024scenicnlgeneratingprobabilisticscenario} relies on the implementation of various prompting techniques that guide an LLM in assembling Scenic scripts. The compound prompting strategy, introduced by ScenicNL, combines CoT reasoning, role-playing as domain experts, and constrained decoding via DSL compiler in a loop, to produce scripts given textual reports. ScenicNL has demonstrated its ability to handle the probabilistic nature and uncertainty of the complex accident descriptions by having the LLM reason about the scenario in different stages, and producing Scenic scripts that capture the probabilistic variations. Implementing iterative compilability (but not executability) checks, ScenicNL ensures higher scripts syntactical quality. 
    
    Talk2Traffic~\cite{Sheng_2025_CVPR} adapts and extends examples from the ChatScene and Scenic repositories, passes them along with the text descriptions of scenarios to LLM, and generates scripts in Scenic format. The method leverages incorporation of human feedback for further improvement of generated scripts. The authors evaluate results using metrics established in ChatScene and SafeBench~\cite{xu2022safebenchbenchmarkingplatformsafety}, also reporting good executability results, however the definition of this metric as well as implementation details remain unclear for now due to non-availability of the source code.  
    
    The reviewed works highlight the potential of LLMs for generating Scenic scripts, however, several approaches operate in a single-pass fashion, which often leads to faulty outputs when scenarios fall outside their training or retrieval domains. Moreover, most methods do not automatically validate whether the generated scripts can be successfully executed in simulation software, thereby overlooking potential runtime or simulation errors that may occur despite syntactic correctness. Therefore, in this work, we address the aforementioned limitations by introducing a set of contributions that enable fully automatic scenario generation from text descriptions. Our approach incorporates multi-stage validation, including final rendering test in the CARLA simulator, to ensure 
    actual executability in the simulation environment, while preserving semantical conformity between generated scripts and initial descriptions.
    
	\section{Methodology}
	\label{sec:methodology}
\begin{figure*}[t]
      \centering
       \includegraphics[width=0.95\linewidth]{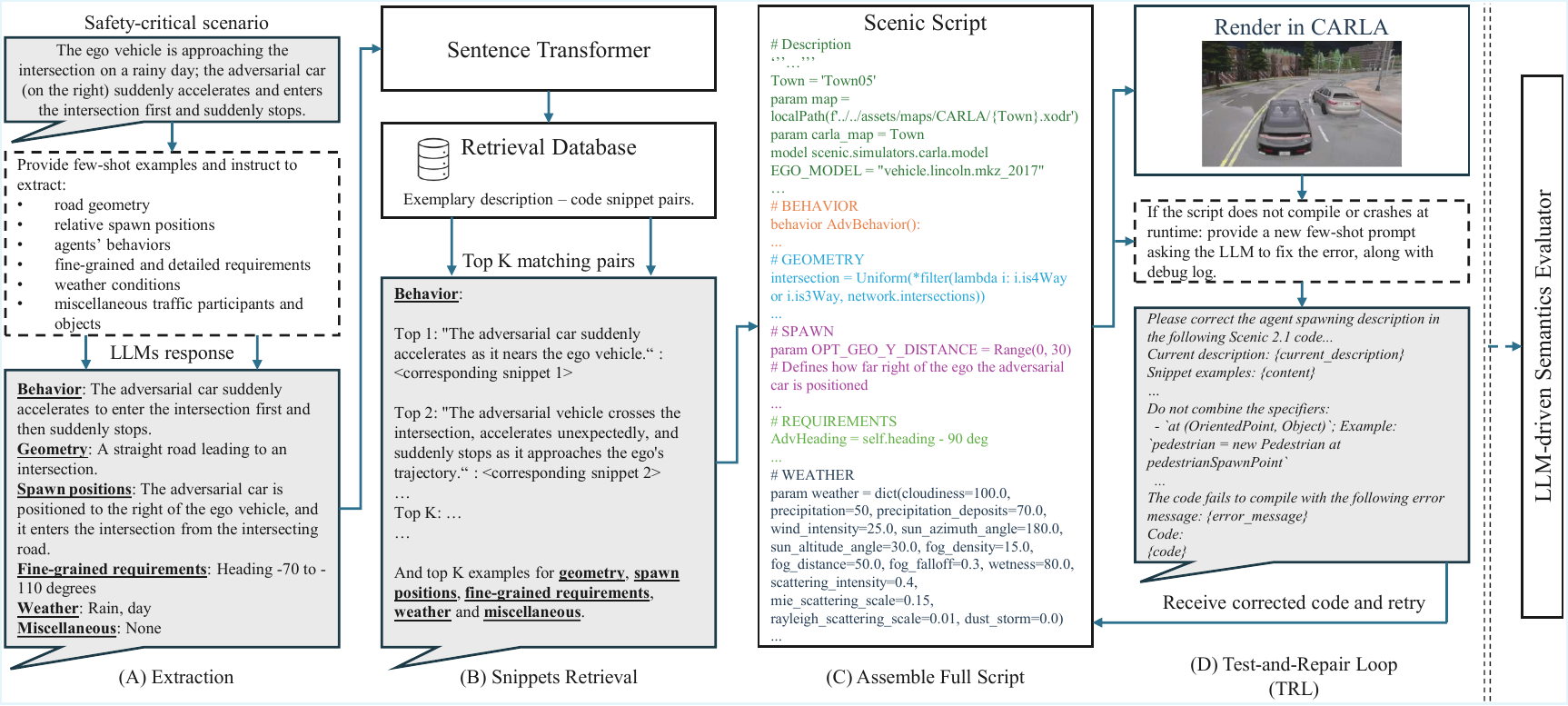}
    
       \caption{\method pipeline (modules A-D) and LLM-driven semantics evaluator.}
       \label{fig:teaser}
    \end{figure*}

	The \method pipeline builds upon the architecture and codebase of ChatScene~\cite{zhang2024chatscene}, which serves as our baseline. An overview of the \method framework is shown in Figure~\ref{fig:teaser}. While it retains the same multi-stage structure for core script generation, \method introduces a key enhancement: an iterative Test-and-Repair Loop (TRL) (cf. Figure~\ref{fig:teaser}, module \textbf{D}). Additionally, we introduce several improvements to modules \textbf{A–C} (cf. Figure~\ref{fig:teaser}). Specifically, we augment the semantical snippet extraction process by incorporating fine-grained sub-components, including: detailed requirements, various objects and traffic participants other than ego and adversary, and weather conditions. Notably, these additions enable more precise information extraction, clearer semantic separation across scenario elements, and a reduced processing burden on other semantical components. Furthermore, we revise and expand the knowledge base of description–snippet pairs (cf. Figure~\ref{fig:teaser}, module \textbf{B}). This includes correcting prior errors, adapting the format to support Scenic 3.0, 
    and introducing new entries to cover underrepresented situations, such as highway driving, bicycles, and roadside-parked vehicles.
    
    By default, \method uses OpenAI’s GPT-4o~\cite{openai2024gpt4technicalreport} for extraction, assembly, and repair. It also supports alternative LLMs such as Google’s Gemini 2.0 Flash~\cite{geminiteam2025geminifamilyhighlycapable} and DeepSeek-V3~\cite{deepseekai2025deepseekv3technicalreport}, selected for their consistently high performance on the LM-Arena leaderboard\footnote[3]{\url{https://lmarena.ai/leaderboard}}~\cite{chen2024llmarenaassessingcapabilitieslarge} as of submission date. 
    
    The remainder of the section provides a detailed description of the overall pipeline, starting with the natural language traffic scenario descriptions and followed by an in-depth explanation of modules A-D. 
    
    
    
    \subsection{Scenario Prompts}
    In line with the baseline, we use 40 prompts describing diverse safety-critical traffic scenarios (each describing interaction between ego vehicle and adversarial agent, with various levels of other details), grouped into eight categories, depicted in the header of Table~\ref{tab:numerical_eval}. Each scenario is defined through a natural language description
    . A sample of a scenario prompt (starting point) is depicted in Figure~\ref {fig:teaser} (module A, upper left).
     
    \subsection{Module A: Extraction}
    
    The extraction stage follows the approach of ChatScene, with the primary distinction being the extension of semantic sub-categories within scenario descriptions. \textbf{First}, \method processes natural language descriptions input of traffic scenarios. \textbf{Second}, an LLM is prompted using a few-shot template that demonstrates how to identify various semantic components. \textbf{Third,} the LLM then extracts key structural elements.
    
    Building upon the baseline, \method expands the semantic categorization by introducing three additional components: (1) fine-grained requirements and constraints (e.g., specific headings or positional constraints), (2) various objects and traffic participants other than ego and adversarial agent, and (3) weather conditions. These extensions use similar prompting strategies and extraction logic as ChatScene, allowing seamless integration. When relevant textual cues are present in the input description, these new components are automatically extracted. These improvements ensure that critical traffic scenario details, such as precise behavioral constraints, 
    environmental factors, and static or peripheral objects are preserved and incorporated. Although some of these elements may not directly interact with the ego or adversarial agents, they contribute meaningful complexity to the overall traffic scenario.
    
    \subsection{Module B: Retrieval}
    
    In the retrieval stage (cf. Figure~\ref{fig:teaser}, module B), \method mirrors the baseline while retrieving relevant code snippets from the knowledge base for each extracted semantic component (previous step). This retrieval process relies on the Sentence-T5 transformer~\cite{ni-etal-2022-sentence} for sentence embedding.
    
    Our improvements focus on refining and expanding the knowledge base, initially sourced in the baseline from the Scenic repository\footnote[4]{\url{https://github.com/BerkeleyLearnVerify/Scenic/tree/main/examples/carla}}. We correct and update existing snippets to comply with the Scenic 3.0 syntax, thereby enhancing retrieval accuracy and supporting seamless script assembly. In addition, \method incorporates new code snippets corresponding to the extended semantic categories. To accommodate these additions, both the retrieval process and knowledge base have been adapted accordingly.
    \method adopts top-k retrieval with a default setting of $k = 2$. In contrast, the baseline defaults to $k = 1$ and skips interaction with LLM in this case, directly using the snippets from the knowledge base in the full code assembly. In \method, upon retrieving the relevant description–snippet pairs, the LLM, guided by a few-shot prompt, generates new or refined code snippets.
    
    \begin{figure*}[ht!]
      \centering
       \includegraphics[width=0.9\linewidth]{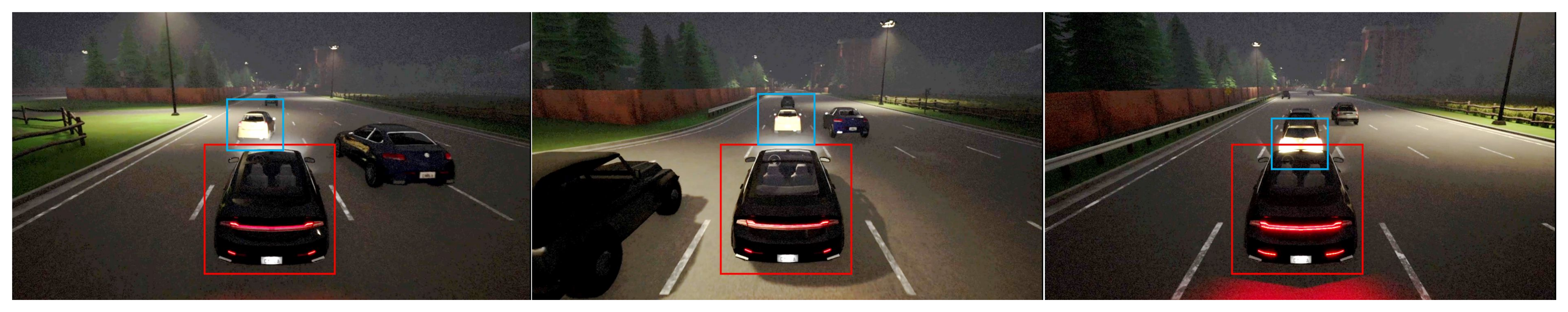}
    
       \caption{A sequence of scenario events: during nighttime with light fog, the \textbf{adversarial vehicle} (blue rectangle) performs a lane change, merging into the same lane as the \textbf{ego vehicle} (red rectangle), and subsequently brakes in front of an obstacle.}
    
       \label{fig:screenshot_sequence}
    \end{figure*}
    \subsection{Module C: Script Assembly}
    
    Following the retrieval process (module B), individual snippets are assembled into a complete Scenic script
    , combining both retrieved and default components, with clearly separated semantic regions. Once the complete Scenic script is assembled, \method invokes the Scenic compiler to generate a Scenario object from the script file. During this compilation phase (transition step between modules C and D), errors such as syntax issues, undefined variables, or missing language constructs (often caused by mismatched or incomplete sub-snippets) are detected and logged. However, successful compilation alone does not guarantee functional correctness, as additional issues can arise during the simulation phase, such as spawn failures or runtime errors. Therefore, upon passing the compilability check, the complete Scenic scripts’ executability is also tested by running them in the CARLA simulator.

    \subsection{Module D: Test-and-Repair Loop (TRL)}

    If compilation and execution tests fail, then the script is passed on to the Test-and-Repair Loop (TRL) (cf. Figure \ref{fig:teaser}, module D) for further post-processing.
    \method differs significantly from the baseline by introducing the automated, iterative TRL
    . This module constitutes the core novelty of our approach. While the baseline follows a single-pass pipeline, producing Scenic scripts without validation and requiring manual debugging in case of errors, \method aims to enhance automation by iteratively repairing scripts to produce fully functional outputs with no user involvement.
     
    In each iteration, structured diagnostic feedback, including error messages, stack traces, and debug information, is passed to an LLM via a carefully crafted automated few-shot prompt. This prompt includes the original scenario description, the most recent script version, and detailed instructions along with selected examples and code snippets specific to the TRL stage. The LLM is tasked with identifying and correcting the errors of the script based on this feedback. The revised script is then recompiled and re-tested in simulation. This process repeats until either the script passes both stages successfully or a predefined maximum number of iterations (set by a user) is reached.
    If the script passes both compilation and simulation steps, it is saved, and relevant statistics are recorded. If all repair attempts fail, the final version of the script is still saved, allowing for manual correction. Furthermore, \method then proceeds to the next execution run or next scenario.
    
    Importantly, \method transforms the generation pipeline into an autonomous and adaptive debugging agent, acting as a mediator between our tool, the simulator and the LLM. This results in significantly improved robustness and execution success rates compared to baseline. For demonstration purposes, in Figure~\ref{fig:screenshot_sequence}, we present a sequence of frames illustrating key events from a safety-critical scenario rendered in the CARLA.
	
	\section{Evaluations}
	\label{sec:evaluations}
	We conducted experiments using 40 natural language scenario descriptions, consistent with the baseline setup. These descriptions are grouped into 8 categories (scenario types), with 5 distinct prompts per category
    . Given the inherent non-determinism of LLMs, and to improve the robustness of the evaluation, we executed each method multiple times per scenario (typically 50, and in some cases, 100 runs - see Table~\ref{tab:numerical_eval}). All experiments were carried out using OpenAI’s GPT-4o model. The maximum number of iterations for TRL was set to 10, and the Scenic simulation environment was allowed up to 15 spawn attempts per script. Since the baseline bypasses LLM-based refinement when the top-k retrieval parameter is set to 1 (default), we set top-k = 2 for all evaluated methods to ensure consistency and comparability.
    
    The first part of the quantitative evaluations focuses on comparing the baseline to enhanced versions of the \method (cf. Table~\ref{tab:numerical_eval}). 
    The second part 
    compares GPT-4o, Gemini 2.0 Flash, and DeepSeek-V3 using the first scenario from each of 8 categories, with 50 runs per scenario.
    
    The qualitative (semantical) evaluation (cf. Tables~\ref{tab:semantics_eval}, \ref{tab:llm_eval_quality}) analyses correspondence between the input prompts and generated scripts. Inspired by OmniTester~\cite{l:24}, we employ an LLM-based evaluator. Additionally, we assess whether it assigns consistent scores to the same script across runs, and similar scores to semantically similar scripts.

    \begin{table*}[ht!]
    \centering\footnotesize
        \caption{Comparison of ESR and RCR for the ChatScene (baseline), ChatScene with TRL (modified), and \method across temperature settings (Temp.) and execution modes (Pass). RCR is omitted for ChatScene (no TRL). "*" and "**" denote results with full or partial 0.01 step size, respectively; "\textsuperscript{+}" marks RCR from successful runs only ("N/A" if all fail).}
    \label{tab:numerical_eval}
    \resizebox{\textwidth}{!}{
        \setlength{\tabcolsep}{3pt}
        \begin{tabular}{clcclccccccccc}
        \toprule
        \textbf{Metric}
          & \textbf{Method}
          & \textbf{TRL}
          & \textbf{Temp.}
          & \textbf{Pass}
          & \cellcolor{gray!20}\scriptsize\makecell{Straight\\Obstacle}
          & \cellcolor{gray!20}\scriptsize\makecell{Turning\\Obstacle}
          & \cellcolor{gray!20}\scriptsize\makecell{Lane\\Changing}
          & \cellcolor{gray!20}\scriptsize\makecell{Vehicle\\Passing}
          & \cellcolor{gray!20}\scriptsize\makecell{Red‑light\\Running}
          & \cellcolor{gray!20}\scriptsize\makecell{Unprotected\\Left‑turn}
          & \cellcolor{gray!20}\scriptsize\makecell{Right\\Turn}
          & \cellcolor{gray!20}\scriptsize\makecell{Crossing\\Negotiation}
          & \textbf{Average} $\uparrow$\\
        \midrule
        \multirow{6}{*}{\textbf{ESR}}
          & ChatScene (baseline)           & \xmark & 1.0     & single        & 0.050*  & 0.002* & 0.000* & 0.000* & 0.004* & 0.026* & 0.020* & 0.008* & 0.014\\
          & ChatScene (modified)         & \cmark & 1.0     & total        & 0.644* & 0.444* & 0.084** & 0.244 & 0.140 & 0.024 & 0.000 & 0.060 & 0.205\\
          & \method             & \cmark & 1.0   & single  & 0.200 & 0.130 & 0.010 & 0.060 & 0.100 & 0.300 & 0.080 & 0.060 & 0.118\\
          & \method            & \cmark & 1.0   & total        & 0.884 & 0.804 & 0.252 & 0.336 & 0.812 & 0.796 & 0.784 & 0.600 & 0.659\\
          & \method            & \cmark & 0.3   & single  & 0.144 & 0.292 & 0.016 & 0.032 & 0.080 & 0.228 & 0.072 & 0.044 & 0.114\\
          & \method            & \cmark & 0.3   & total        & \textbf{0.964} & \textbf{0.872} & \textbf{0.424} & \textbf{0.632} & \textbf{0.892} & \textbf{0.848} & \textbf{0.876} & \textbf{0.728} & \cellcolor{gray!20}\textbf{0.780}\\
        \midrule
        \multirow{3}{*}{\textbf{RCR}}
          & ChatScene (modified)          & \cmark & 1.0     & total        & 3.235* & 4.982* & 4.402**\textsuperscript{+} & 4.503 & 5.623 & 2.875\textsuperscript{+} & N/A & 5.384\textsuperscript{+} & 4.429\\
          & \method            & \cmark & 1.0   & total        & 2.073 & 3.261 & 4.880 & 4.563 & 2.997 & 3.353 & 3.520 & 3.602 & 3.531\\
          & \method            & \cmark & 0.3   & total        & \textbf{1.649} & \textbf{2.262} & \textbf{3.661} & \textbf{2.508} & \textbf{1.927} & \textbf{1.964} & \textbf{2.257} & \textbf{3.134} & \textbf{2.420}\\
        \bottomrule
        \end{tabular}
        }
    \end{table*}
    
    \subsection{Metrics}
    
    Since \method aims to reduce the need for manual intervention in producing executable traffic scenario scripts, we introduce two evaluation metrics: Execution Success Rate (ESR) and Repair Convergence Rate (RCR). ESR measures the proportion of Scenic scripts that are not only syntactically valid and compilable, but also executable in CARLA. The ESR is computed as:
    
    \begin{equation}
        ESR = \frac{N_{\text{success}}}{N_{\text{total}}}    
        \label{eq:esr}
    \end{equation}
    
    \noindent
    where \( N_{\text{total}} \) denotes the total number of generated scripts, and \( N_{\text{success}} \) represents the fraction that successfully compiled and executed without errors.
    
    The RCR metric measures the average number of repair iterations of the TRL. It considers only the runs where TRL led to a successful outcome, excluding cases where the script succeeded on the first attempt or failed entirely after exhausting all repair attempts. RCR is computed as:
    \begin{equation}
    RCR = \frac{1}{N_{\text{repair}}} \sum_{i=1}^{N_{\text{repair}}} A_i
    \label{eq:rcr}
    \end{equation}
    \noindent
    where \( N_{\text{repair}} \) is the number of runs in which the script failed initially, but succeeded after one or more repair attempts, and $A_i$ denotes the total number of repair iterations performed in such run $i$, including the final successful attempt.
    
    Additionally, we introduce Semantic Conformity Score (SCS) for rating correspondence between the initial scenario prompt and the generated Scenic script:
    \begin{equation}
    SCS = \frac{1}{inScore_{\textbf{total}} \times N_{\text{criteria}}} \sum_{i=1}^{N_{\text{criteria}}} Score_{\text{i}}
    \label{eq:scs}
    \end{equation}
    \noindent
    where \( inScore_{\textbf{total}} \) denotes the maximum possible score a script can receive for each scoring criterion (fixed at 10 points in our experiment), and \( N_{\text{criteria}} \) is the total number of criteria used for evaluation (7 in our case), and $Score_i$ represents the actual score obtained for the $i$-th criterion. 
    
    \subsection{Quantitative Evaluations}
    
    \method significantly outperforms (average ESR of 78\%) both the baseline (cf. Table~\ref{tab:numerical_eval}), which often failed to produce valid simulations in several scenario categories, resulting in ESR values close to zero (average of 1.4\%), and the baseline that we extended with TRL (modified, average of 20.5\%). Note: the RCR metric is not applicable to the baseline since it operates in a single-pass script generation mode. We have implemented the latter as an ablation to highlight the isolated impact of the individual improvements that we have introduced in \method. These results highlight the effectiveness of combining TRL with an expanded knowledge base and fine-grained semantic prompting, significantly reducing the need for manual involvement to produce syntactically correct and executable scripts across all evaluated traffic categories. Further improvements are observed when lowering the LLM temperature to 0.3, which yields better ESR and RCR scores compared to both the baseline and \method with the default temperature of 1.0. Lowering the temperature increases determinism, while still allowing for limited output diversity.
    A more comprehensive exploration of temperature and other fine-tuning parameters is left for future work.
    For completeness, we additionally report ChatScene’s one-shot (pre-TRL) results to benchmark the pure baseline. While the overall ESR values are nearly identical (0.014 for ChatScene and 0.012 for ChatScene + TRL), performance varied across scenario categories. These differences can be attributed to the inherent non-determinism of LLM outputs.
    RCR results indicate steady improvement across the methods. Notably, \method's overall success rate was over an order of magnitude higher than the baseline, and significantly higher than the baseline with TRL. This improvement supports our claim of the effectiveness of the proposed combined approach with TRL, more fine-granular extraction prompts, and an extended (and revised) knowledge base. Moreover, RCR numbers show that \method, on average, requires a relatively small number of repair cycles per scenario. Most baseline scenarios were automatically converted into functional Scenic scripts, advancing our goal of a fully automated traffic scenario generation pipeline and its importance for the community.

\begin{table}[ht!]
\centering\footnotesize
\setlength{\tabcolsep}{3pt}
\renewcommand{\arraystretch}{1.1}
\caption{Comparison of Semantical Conformity Scores (SCS) across models and temperature settings.}
\begin{tabular}{lcccc}
\toprule
\textbf{Metric} &
\makecell[c]{\textbf{GPT-4o}\\Temp.=1.0} &
\makecell[c]{\textbf{GPT-4o}\\Temp.=0.3} &
\makecell[c]{\textbf{Gemini 2.0 Flash}\\Temp.=0.3} &
\makecell[c]{\textbf{DeepSeek-V3}\\Temp.=0.3} \\
\midrule
Min. SCS & 64.64\% & \cellcolor{gray!20}\textbf{73.93\%} & 70.36\% & 65.54\% \\
Max. SCS & 99.11\% & \cellcolor{gray!20}\textbf{99.29\%} & 97.32\% & 98.75\% \\
Avg. SCS & 86.55\% & \cellcolor{gray!20}\textbf{88.20\%} & 84.90\% & 85.66\% \\
Std. Dev. (pp) & 11.43 & \cellcolor{gray!20}\textbf{9.07} & 9.54 & 11.05 \\
\bottomrule
\end{tabular}

\label{tab:semantics_eval}
\end{table}

    \begin{table}[ht!]
    \caption{Comparison of SCS ranges and standard deviation for repeated evaluation of the same script in each scenario type.}
    \centering\small
    \setlength{\tabcolsep}{3pt}
    \begin{tabular}{lcccc}
    \toprule
    \makecell[l]{\textbf{Metric}\\Base Scenario} 
      & \makecell{\textbf{Min. SCS}\\in \%}
      & \makecell{\textbf{Max. SCS}\\in \%}
      & \makecell{\textbf{Avg. SCS}\\in \%}
      & \makecell{\textbf{Std. Dev.}\\in pp} \\
    \midrule
    \scriptsize\makecell[l]{Straight Obstacle}       & 94.29 & 100.00 & 98.86 & 2.11 \\
    \scriptsize\makecell[l]{Turning Obstacle}        & 88.57 & 92.86  & 92.00 & 1.54 \\
    \scriptsize\makecell[l]{Lane Changing}           & 84.29 & 100.00 & 93.57 & 6.22 \\
    \scriptsize\makecell[l]{Vehicle Passing}         & 82.86 & 100.00 & 91.29 & 5.73 \\
    \scriptsize\makecell[l]{Red-light Running}       & 81.43 & 100.00 & 94.00 & 6.66 \\
    \scriptsize\makecell[l]{Unprotected Left-turn}   & 94.29 & 100.00 & 97.57 & 2.43 \\
    \scriptsize\makecell[l]{Right Turn}              & 100.00 & 100.00 & 100.00 & 0.00 \\
    \scriptsize\makecell[l]{Crossing Negotiation}    & 82.86 & 100.00 & 97.57 & 5.30 \\
    \textbf{Average}                                                  & \textbf{88.57} & \textbf{99.11}  & \textbf{95.61} & \textbf{3.75} \\
    \bottomrule
    \end{tabular}

    \label{tab:llm_eval_quality}
\end{table}

\begin{table}[t]
\footnotesize
\centering
  \caption{Average Safebench test results conducted using SAC vehicle. \collisionrateabbr: \collisionrate, \runredlightabbr: \runredlight, \runstopsignabbr: \runstopsign, \outofroadabbr: \outofroad, \followrouteabbr: \followroute, \routecompletionabbr: \routecompletion, \timespentabbr: \timespent, \accelerationabbr: \acceleration, \yawvelocityabbr: \yawvelocity, \laneinvasionabbr: \laneinvasion, \overallscoreabbr: \overallscore, $\uparrow$/$\downarrow$: higher/lower the better.}
\setlength{\tabcolsep}{2.5pt}
\renewcommand{\arraystretch}{0.95}
\begin{tabular}{lcccccc}
\toprule
Metric \textbf{Methods} & \textbf{LC} & \textbf{AS} & \textbf{CS} & \textbf{AT} & \textbf{ChatScene} & \textbf{\method (ours)} \\
\midrule
\multicolumn{7}{c}{\cellcolor{gray!20}\textbf{Safety Level}} \\
CR $\uparrow$ & 0.584 & 0.586 & 0.676 & 0.627 & \textbf{0.831} & 0.564 \\
RR $\uparrow$ & 0.326 & 0.300 & 0.313 & 0.312 & 0.179 & \textbf{0.359} \\
SS $\uparrow$ & 0.158 & 0.160 & \textbf{0.161} & 0.158 & 0.143 & 0.090 \\
OR $\uparrow$ & 0.032 & 0.025 & \textbf{0.036} & 0.028 & 0.035 & 0.033 \\
\midrule
\multicolumn{7}{c}{\cellcolor{gray!20}\textbf{Functionality Level}} \\
RF $\downarrow$ & 0.894 & 0.891 & 0.890 & 0.893 & 0.833 & \textbf{0.753} \\
Comp $\downarrow$ & 0.731 & 0.745 & 0.741 & 0.726 & 0.544 & \textbf{0.440} \\
TS $\uparrow$ & 0.216 & 0.261 & 0.244 & \textbf{0.279} & 0.223 & 0.125 \\
\midrule
\multicolumn{7}{c}{\cellcolor{gray!20}\textbf{Etiquette Level}} \\
ACC $\uparrow$ & 0.211 & 0.203 & 0.215 & 0.219 & \textbf{0.705} & 0.359 \\
YV $\uparrow$ & 0.243 & 0.245 & 0.243 & 0.248 & \textbf{0.532} & 0.226 \\
LI $\uparrow$ & 0.112 & 0.127 & 0.131 & 0.137 & \textbf{0.243} & 0.052 \\
\midrule
OS $\downarrow$ & 0.619 & 0.620 & \textbf{0.573} & 0.596 & \textbf{0.482} & 0.614 \\
\bottomrule
\end{tabular}
\label{tab:diagnostic_singlecol}
\end{table}

   
    \subsection{Qualitative Evaluations} 
    Inspired by OmniTester’s LLM-based evaluator, we introduce our own module to assess the semantic correspondence between scenario descriptions and generated Scenic scripts using Equation~\ref{eq:scs}. The evaluator, implemented with Google Gemini 2.5 Flash, operates in two stages: it is first primed with a condensed Scenic reference and curated examples, then evaluates description–script pairs using a structured few-shot prompt with seven scoring criteria (adversarial type, behavior, geometry, weather, elements, spawn, and requirements). We conduct two evaluations: inter-scenario/inter-LLM SCS comparison and scoring consistency analysis (cf. Table \ref{tab:semantics_eval}).
    
    \noindent
    \textbf{Semantical Evaluation 1.}
    In the first experiment, we selected the first 10 generated scripts per scenario type for each LLM, using the same criteria as in the inter-LLM ESR and RCR comparison (i.e., the first scenario from each category) and evaluated them using our LLM-based evaluator.
    The key results are reported in Table~\ref{tab:semantics_eval}
    . The results demonstrate that the inter-LLM, inter-scenario-type average SCS varying between 84.9\% (Gemini 2.0 Flash) and 88.2\% (GPT-4o, temperature=0.3), while the lowest inter-scenario score lies at 50\% (GPT-4o, temperature 1.0, Lane Changing; Gemini 2.0 Flash, temperature=0.3, Red-Light Running)
    .
    Showing such a high average SCS further improves our confidence in \method's ability to preserve the semantical conformity between the initial scenario description and the resulting Scenic scripts.
    \\
    \textbf{Semantical Evaluation 2.} 
    To assess the soundness of our approach, we analyze the standard deviation of SCS in two settings (cf. Table~\ref{tab:semantics_eval}): repeated scoring of the same script (10 times each across 8 scenario types), and scoring across 8 groups of 10 different scripts (one group per scenario type). This allows us to evaluate the consistency and reliability of the LLM-based evaluator, expecting similar scores for the same and semantically similar scripts. 
    We exclude other LLMs here, as the goal is to evaluate the stability of the evaluator itself, and including additional models would provide limited added value. Table~\ref{tab:llm_eval_quality} shows a low average standard deviation of 3.75 percentage points when the same script is evaluated repeatedly across scenario categories. In contrast, Table~\ref{tab:semantics_eval} presents higher standard deviations in the inter-scenario, inter-LLM comparison involving different scripts (10 per scenario), which is expected due to script variability. These results indicate that the LLM-based semantic evaluator produces consistent scores, supporting the reliability of the scoring approach.

    \subsection{Criticality Evaluations}

    We used SafeBench~\cite{xu2022safebenchbenchmarkingplatformsafety} to confirm compatibility of \method with downstream application of the ChatScene. Here, we used an ego vehicle trained with the Soft Actor-Critic (SAC)~\cite{haarnoja18b} deep RL methodology provided by SafeBench. The other two trained ego agents, reported in ChatScene, were not provided / published by authors, and therefore were excluded from this evaluation. After an increase of criticality of our scenarios (5 per each of the 8 base scenario categories) using SafeBench, we compared the performance of the ego vehicle on our scenarios across three levels (cf. Table~\ref{tab:diagnostic_singlecol}) to the ChatScene and its baseline adversary- and knowledge-based scenario generation techniques: Learning-to-collide (LC)~\cite{ding2020learning}, AdvSim (AS)~\cite{wang2021advsim}, Carla Scenario Generator (CS)~\cite{scenariorunner} and Adversarial Trajectory Optimization (AT)~\cite{zhang2022adversarial}. The results show general performance alignment of the \method with the baselines, even outperforming them in several metrics (frequency of running red lights (RR), route following stability (RF) and average percentage of route completion (Comp), while achieving comparable scores on others, reflecting trade-offs in the behavior of the generated scenarios. These metrics are computed within SafeBench, which is configured to deliberately increase scenario criticality for stress-testing; \method is focused on automated scenario generation rather than on optimizing this particular downstream task, which could be a focus of a follow-up work. It is important to note that in ChatScene, authors manually polished generated scripts, introducing bias into the generation process, while also increasing criticality of events. Our evaluation showed ChatScene achieved ESR of only 1.4\% for automatically generated (no manual involvement) scenarios. In contrast, ARISE operates with minimal human intervention, achieving ESR of 78\%. 
    Importantly, in our experiments we found that some of the criticality metrics reported in ChatScene appear 
    unrealistic. For instance, the reported average acceleration (ACC) of 0.705 (cf. Table~\ref{tab:diagnostic_singlecol}), in combination with virtually guaranteed collision chance, would correspond to implausibly aggressive vehicle dynamics.

    
	\section{Future Work and Discussion}
	\label{sec:futurework}
	As a future enhancement, we plan to integrate the LLM-based evaluator into the TRL as a stricter filter for output scripts. Additionally, allowing users to adjust the weights of scoring criteria would enable finer control, emphasizing specific aspects of script quality. Further improvements could come from experimenting with different LLMs and parameters (e.g., temperature). Since \method generates Scenic scripts, future work will explore Scenic-to-OpenSCENARIO translation (to align with the open industry-required formats), expanding input types (e.g., accident reports, non-verbal data), and benchmarking against a wider range of tools.
    The main objective is to maximize the automation of traffic scenario generation while minimizing human intervention. Accordingly, we focus our downstream evaluation on criticality, primarily to demonstrate that \method can also support this stage of the pipeline; however, the impact of generated scenarios on downstream criticality (e.g., SafeBench) is left for future work.

	\section{Conclusion}
	\label{sec:conclusion}
	We introduce \method, an automated tool that converts natural language traffic descriptions into Scenic simulation scripts via an adaptive, feedback-driven process. By incorporating a refinement loop, \method effectively translates free-form scenario descriptions into executable code while preserving their semantics. Our experiments show that \method achieves significantly higher execution success rates than the baseline, setting a new benchmark by producing both functionally and semantically valid scripts for most inputs. Leveraging a knowledge-augmented LLM, a test-and-repair loop, and improved prompt and knowledge base design, \method ensures scripts are syntactically correct and require little to no user intervention. This demonstrates the feasibility of fully automated traffic scenario generation for autonomous driving.

\section{Acknowledgment}
This work was supported by the European Union’s Horizon Europe research and innovation programme under grant agreement No. 101076360 and in part by the European Regional Development Fund (ERDF) under grant agreement No. EFRE-AuF-0000866.
    
	
	\bibliographystyle{IEEEtran}
	\bibliography{root} 
	
\end{document}